# Soliton modes, stability, and drift in optical lattices with spatially modulated nonlinearity

Yaroslav V. Kartashov, Victor A. Vysloukh, and Lluis Torner

*ICFO-Institut de Ciencies Fotoniques, and Universitat Politecnica de Catalunya, Mediterranean Technology Park, 08860 Castelldefels (Barcelona), Spain*

We put forward new properties of lattice solitons in materials and geometries where both, the linear refractive index and the nonlinearity are spatially modulated. We show that the interplay between linear and out-of-phase nonlinear refractive index modulations results in new soliton properties, including modifications of the soliton stability and transverse mobility, as well as shape transformations that may be controlled, e.g., by varying the light intensity.

Periodic optical materials offer a unique landscape to control the propagation of light. In the presence of nonlinearity such materials give rise to lattice solitons [1-3]. The current technological state-of-the-art allows relatively deep simultaneous modulation of the linear refractive index and the nonlinearity coefficient of the material [4-6]. This is the case, e.g., of arrays made by Ti-indiffusion in $LiNbO_3$ crystals [4], where nonlinearity depends on the concentration of dopants, and also of arrays written in glass by high intensity fs laser pulses where optical damage results in an increase of the refractive index accompanied by a significant decrease of the nonlinear coefficient [5,6]. Thus, one can fabricate lattices with out-of-phase modulations of refractive index and nonlinearity. Such lattices have been studied using discrete models [7,8] and an enhancement of soliton mobility has been predicted. Similar phenomena arise in Bose-Einstein condensates [9-13]. In particular, the conditions for mutual cancellation of effects of shallow linear and nonlinear lattices have been addressed using perturbative approaches [13].

In this Letter we address the structural transformations of shapes, drastic modification of stability properties, and transverse mobility for solitons in continuous lattices where the nonlinearity, diffraction and refractive index modulation compete on similar footing. In such lattices high-power odd solitons may develop two-humped intensity distributions and destabilize, while even solitons fuse into a single stable peak located between lattice channels. Drift destabilization of odd and even solitons leads to an enhanced mobility.

We describe propagation of light beam in the lattice with an out-of-phase modulation of both the linear refractive index and the nonlinearity with nonlinear Schrödinger equation for the field amplitude $q$. Namely,

$$i\frac{\partial q}{\partial \xi} = -\frac{1}{2}\frac{\partial^2 q}{\partial \eta^2} - [1 - \sigma R(\eta)]|q|^2 q - pR(\eta)q. \qquad (1)$$

Here $\eta$ and $\xi$ are normalized transverse and longitudinal coordinates, respectively; $p$ and $\sigma$ are the depths of modulation of the refractive index and nonlinearity, while $R(\eta) = \cos^2(\Omega \eta)$ describes the linear lattice profile. The nonlinear coefficient $\gamma = 1 - \sigma R$ attains minima at the points where the refractive index has a maximum, as it occurs upon laser writing of waveguide arrays [6]. We are interested in stationary solutions $q = w(\eta)\exp(ib\xi)$ of Eq. (1) that can be characterized by the propagation constant $b$ and the energy flow $U = \int_{-\infty}^{\infty} |q|^2 \, d\eta$. Substitution of perturbed solutions of the form $q = [w(\eta) + u(\eta)\exp(\delta\xi) + iv(\eta)\exp(\delta\xi)]\exp(ib\xi)$, where $u, v \ll w$, into Eq. (1) and linearization yields the eigenvalue problem $\delta u = -(1/2)d^2v/d\eta^2 + bv - (1 - \sigma R)vw^2 - pRv$, $\delta v = (1/2)d^2u/d\eta^2 - bu + 3(1 - \sigma R)uw^2 + pRu$, that we solved numerically to obtain the growth rate $\delta = \delta_r + i\delta_i$. We consider here sufficiently deep linear lattices ($p = 4$ and $\Omega = 2$) that result in notable shape modulations of solitons. We vary $\sigma$ and $U$ to study their impact on the soliton shape, stability and mobility.

Lattices with modulated nonlinearity support odd, even, dipole, and triple-mode solitons [Fig. 1]. Odd solitons bifurcate from Bloch waves (i.e. they exist for $b \geq b_{cl}$, where $b_{cl}$ is the lower edge of the semi-infinite gap) and at low powers they exhibit single intensity maximum. The power of such soliton increases monotonically with $b$ and the intensity maximum remains in the same channel only for $\sigma \leq 0.8$. For $\sigma > 0.8$, the spatially non-uniform self-focusing dominates over the linear refraction, resulting in the light deflection toward the regions where the nonlinearity is stronger. This is accompanied by a reshaping of the single-peak odd soliton into two peaks located around the minima of $R$ [Fig. 1(b)], while dependence $U(b)$ becomes nonmonotonic [Fig. 2(a)]. Even solitons exhibiting two intensity maxima may fuse into a single peak located between the maxima of $R$ for high enough $\sigma$ [Fig. 1(d)], which also results in nonmonotonic $U(b)$ dependence [Fig. 2(b)]. With growth of $\sigma$ such reshaping transformation occurs at smaller $b$ values. Dipole and triple-mode solitons at $\sigma \neq 0$ feature both lower $b_{\text{cl}}$ and upper $b_{\text{cu}}$ cutoffs [Figs. 1(e) and 1(f)]. The derivative $dU/db$ diverges at both cutoffs $b = b_{\text{cl}}, b_{\text{cu}}$. At $b = b_{\text{cu}}$ the humps of dipole solitons shift to the minima of linear lattice. The domain of dipole soliton existence shrinks with $\sigma$ and they cease to exist at $\sigma \geq 0.847$ [Fig. 2(e)]. This is in contrast to usual lattices where such states always exist. The existence domain of triple-mode solitons is even narrower $(\sigma < 0.762)$.

We found that the Vakhitov-Kolokolov criterion fails to predict the stability of odd solitons. Despite the fact that at $\sigma < 0.8$ one has $dU/db > 0$ for the entire odd soliton family, such solitons become unstable above a critical propagation constant $b_{\text{o}}$ [Fig. 2(d)]. This instability is of a drift origin and it occurs because of violation of spectral stability conditions [14]. The drift instability causes a rapid displacement of the soliton center into regions where $R$ attains a local minimum. The width of the stability domain $b_{\text{cl}} \leq b \leq b_{\text{o}}$ for odd solitons decreases with $\sigma$ and then vanishes completely [Fig. 2(c)]. Even solitons become stable for $b \geq b_{\text{e}}$ [Fig. 2(d)], and the width of their stability domain expands with $\sigma$, so that the entire family of even solitons becomes stable for the same $\sigma$ value at which destabilization of odd soliton family occurs. In all cases $b_{\text{e}} > b_{\text{o}}$. Also, there always exists a domain $b_{\text{o}} \leq b \leq b_{\text{e}}$ [shaded region in Fig. 2(c)] where both odd and even solitons are unstable. Odd solitons are stable below this domain, while even solitons are stable above. Multipole-mode solitons can be robust too, although the width of the stability domain [Fig. 2(e)] shrinks with increasing $\sigma$, mainly due to a broadening of the instability region near the upper cutoff [Fig. 2(f)].

Solitons discussed above can be excited by Gaussian beams $A \exp(-\eta^2)$, but excitation scenarios differ remarkably from those in usual lattices. When the beam center coincides with the maximum of $R$ and $\sigma < 0.8$, one excites stable odd soliton at low powers [Fig. 3(a)]. If the power is high enough, one excites unstable odd solitons that leave the input channel and drift across the lattice until they are trapped in the region between maxima of $R$, thereby forming single-hump stable even solitons [Fig. 3(b)]. For higher soliton powers, one may excite two diverging odd solitons walking across the lattice [Fig. 3(c)], while in usual lattices such beams would remain in the input channel. For $\sigma > 0.8$ and high input power original beam may split into two beams oscillating around minima of $R$ [Fig. 3(d)].

We found that soliton mobility is intimately related to their stability. To set soliton in motion across the lattice, one has to impose a phase tilt $\exp(i\alpha\eta)$. In lattices with $\sigma = 0$ [15] the critical tilt $\alpha_{\rm cr}$ at which soliton leaves the input channel grows rapidly and monotonically with power $U$ (note that a similar behavior was encountered in nonlinear lattices [16]). In contrast, for odd solitons of Eq. (1) the critical tilt turns out to be a non-monotonic function of $b$ [Fig. 4(a)], and $\alpha_{\rm cr}$ vanishes at $b = b_{\rm cl}$ as well as at the point $b = b_{\rm o}$ where odd solitons become unstable. Very small tilts result in an almost radiationless motion of odd solitons across the lattice in this region of drift instability ($b > b_{\rm o}$). For tilts only slightly exceeding $\alpha_{\rm cr}$ solitons move across the lattice almost without losses [Fig. 4(c)] even in the stability region ($b < b_{\rm o}$), and they do not experience trapping in subsequent lattice channels as it occurs in lattices with $\sigma = 0$ [15]. Thus, even in relatively deep linear lattices, an out-of-phase periodic nonlinearity can drastically suppress radiative losses (not only for particular ratio of depths of linear and nonlinear lattices [13]), and this suppression is more pronounced at larger powers. The situation is similar for even solitons: they become mobile for $b < b_{\rm e}$, while $\alpha_{\rm cr}$ increases monotonically with $b$ for $b > b_{\rm e}$ [Fig. 4(b)].

In conclusion, the central result of this Letter is that out-of-phase periodic nonlinearity modulation results in a reshaping of lattice solitons and in a substantial modification of the stability properties of the soliton families. Last but not least, the nonlinearity modulation may lead to a remarkable enhancement of the soliton mobility.

# References with titles

# References without titles

# Figure captions

Figure 1.   Profiles of odd solitons with (a) $b=5.5$ and (b) $b=12$ at $\sigma=0.85$. Profiles of even solitons with (c) $b=3.7$ and (d) $b=7$ at $\sigma=0.3$. Profiles of (e) dipole and (f) triple-mode solitons with $b=4$ at $\sigma=0.4$. In shaded regions $R \geq 1/2$, while in white regions $R < 1/2$. Red lines show profile of nonlinearity coefficient $\gamma = 1 - \sigma R$.

Figure 2.   $U$ versus $b$ for (a) odd soliton at $\sigma=0.85$ and (b) even soliton at $\sigma=0.3$. Points marked by circles in (a) correspond to solitons in Figs. 1(a) and 1(b), while points marked by circles in (b) correspond to solitons in Figs. 1(c) and 1(d). (c) Critical $b$ values for stabilization of odd and even solitons versus $\sigma$. In shaded area both odd and even solitons are unstable. (d) $\delta_r$ versus $b$ for odd soliton at $\sigma=0.85$ (curve 1) and even soliton at $\sigma=0.3$ (curve 2). (e) Domains of instability (shaded region) and stability (white region embedded inside shaded one) on $(\sigma,b)$ plane for dipole solitons. (f) $\delta_r$ versus $b$ for dipole soliton at $\sigma=0.4$.

Figure 3.   Dynamics of soliton excitation by Gaussian input beams at (a) $A=2$, $\sigma=0.2$, (b) $A=3.1$, $\sigma=0.6$, (c) $A=5.6$, $\sigma=0.2$, and (d) $A=5$, $\sigma=0.9$.

Figure 4.   Critical angle versus $b$ for (a) odd and (b) even solitons. (c) Propagation dynamics of odd solitons with $b=9.8$ launched in the lattice at two different angles. The field modulus distributions corresponding to different input angles are superimposed. In all cases $\sigma=0.4$.

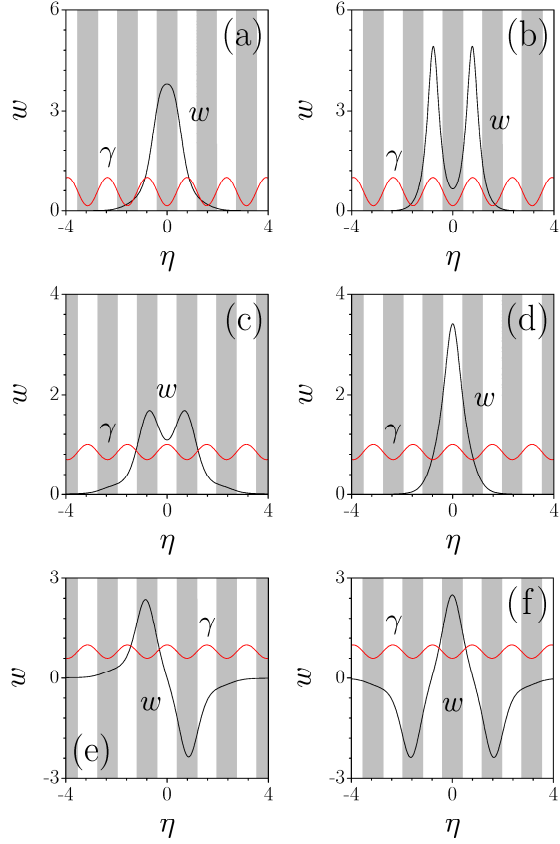

Figure 1. Profiles of odd solitons with (a) $b = 5.5$ and (b) $b = 12$ at $\sigma = 0.85$. Profiles of even solitons with (c) $b = 3.7$ and (d) $b = 7$ at $\sigma = 0.3$. Profiles of (e) dipole and (f) triple-mode solitons with $b = 4$ at $\sigma = 0.4$. In shaded regions $R \geq 1/2$, while in white regions $R < 1/2$. Red lines show profile of nonlinearity coefficient $\gamma = 1 - \sigma R$.

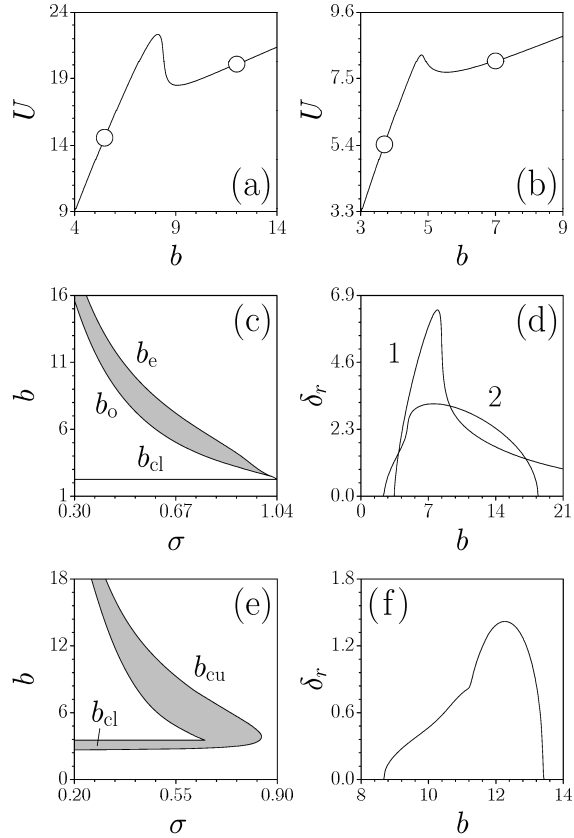

Figure 2. $U$ versus $b$ for (a) odd soliton at $\sigma = 0.85$ and (b) even soliton at $\sigma = 0.3$. Points marked by circles in (a) correspond to solitons in Figs. 1(a) and 1(b), while points marked by circles in (b) correspond to solitons in Figs. 1(c) and 1(d). (c) Critical $b$ values for stabilization of odd and even solitons versus $\sigma$. In shaded area both odd and even solitons are unstable. (d) $\delta_r$ versus $b$ for odd soliton at $\sigma = 0.85$ (curve 1) and even soliton at $\sigma = 0.3$ (curve 2). (e) Domains of instability (shaded region) and stability (white region embedded inside shaded one) on $(\sigma, b)$ plane for dipole solitons. (f) $\delta_r$ versus $b$ for dipole soliton at $\sigma = 0.4$.

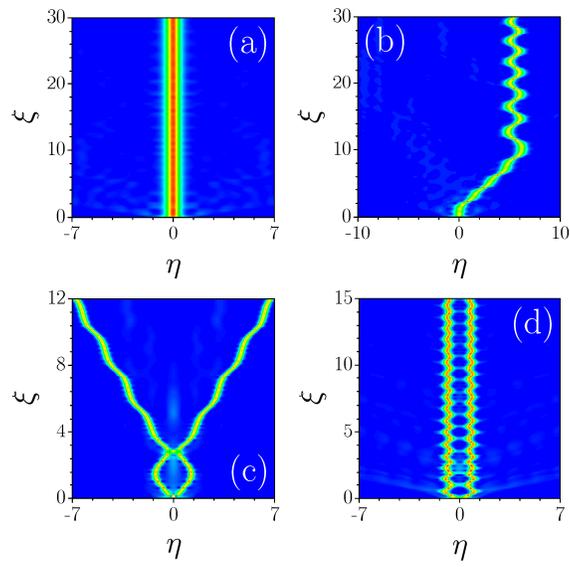

Figure 3. Dynamics of soliton excitation by Gaussian input beams at (a) $A = 2$, $\sigma = 0.2$, (b) $A = 3.1$, $\sigma = 0.6$, (c) $A = 5.6$, $\sigma = 0.2$, and (d) $A = 5$, $\sigma = 0.9$.

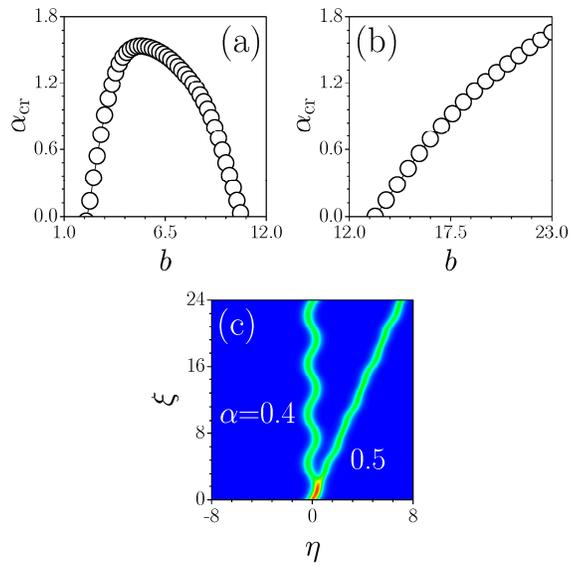

Figure 4.  Critical angle versus $b$ for (a) odd and (b) even solitons. (c) Propagation dynamics of odd solitons with $b = 9.8$ launched in the lattice at two different angles. The field modulus distributions corresponding to different input angles are superimposed. In all cases $\sigma = 0.4$.